\documentclass{iau}

\usepackage{amsmath}
\usepackage{graphicx}
\usepackage{multirow}

\usepackage{etex} 

\begin{document}

\lefttitle{L.I. Gurvits et al.}
\righttitle{A potential distant multimessenger?}

\jnlPage{1}{7}
\jnlDoiYr{2023}
\doival{10.1017/xxxxx}

\aopheadtitle{Proceedings IAU Symposium}
\editors{Y. Liodakis et al., eds.}

\title{J2102+6015: a potential distant multimessenger?}

\author{Leonid~I.~Gurvits$^{1,2}$, S\'{a}ndor~Frey$^{3,4}$, M\'{a}t\'{e}~Krezinger$^{5,3}$, Oleg~Titov$^{6}$, Tao An$^{7}$, Yingkang Zhang$^{7}$, Alexander~G.~Polnarev$^{8}$, Krisztina \'E.  Gab{\'a}nyi$^{5,9,3}$, Krisztina~Perger$^{3}$, Alexey~Melnikov$^{10}$}

\affiliation{$^1$Joint Institute for VLBI ERIC, Oude Hoogeveensedijk 4, 7991~PD Dwingeloo, The Netherlands}
\affiliation{$^2$Aerospace Faculty, Delft University of Technology, Kluyverweg 1, 2629~HS Delft, The~Netherlands}
\affiliation{$^3$Konkoly Observatory, ELKH Research Centre for Astronomy and Earth Sciences \\ (MTA Centre of Excellence), Konkoly Thege Mikl\'{o}s \'{u}t 15-17, H-1121, Budapest, Hungary}
\affiliation{$^4$Institute of Physics and Astronomy, ELTE E\"{o}tv\"{o}s Lor\'{a}nd University, P\'{a}zm\'{a}ny P\'{e}ter s\'{e}t\'{a}ny 1/A, H-1117, Budapest, Hungary}
\affiliation{$^5$Department of Astronomy, Institute of Physics and Astronomy, ELTE E\"{o}tv\"{o}s Lor\'{a}nd University, P\'{a}zm\'{a}ny P\'{e}ter s\'{e}t\'{a}ny 1/A, H-1117, Budapest, Hungary}
\affiliation{$^6$Geoscience Australia, PO Box 378, Canberra 2601, Australia}
\affiliation{$^7$Shanghai Astronomical Observatory, CAS, 80 Nandan Road, Shanghai 200030, China}
\affiliation{$^{8}$Queen Mary University of London, London E1~4NS, United Kingdom}
\affiliation{$^{9}$ELKH-ELTE Extragalactic Astrophysics Research Group, E\"{o}tv\"{o}s Lor\'{a}nd University, \\ P\'{a}zm\'{a}ny P\'{e}ter s\'{e}t\'{a}ny 1/A, H-1117, Budapest, Hungary}
\affiliation{$^{10}$Institute of Applied Astronomy, Russian Academy of Sciences, Kutuzova Embankment 10, \\ St. Petersburg 191187, Russia}

\begin{abstract}
We present and briefly discuss results of several studies of the source J2102$+$6015 with tentatively defined redshift $z=4.575$ which demonstrates unusual properties in imaging and astrometric VLBI observations. Its properties might be considered as indications on the supermassive black hole binary which can be considered as a so far rare example of a high-redshift source of known electromagnetic and, possibly, predictable gravitational wave emissions.
\end{abstract}

\begin{keywords}
SMBH binary, AGN, VLBI, multimessenger
\end{keywords}

\maketitle

\section{Introduction}

Active galactic nuclei (AGN) are the most powerful ``engines'' in the Universe. Due to their enormous luminosity they are visible and can be studied with modern astronomy instruments at any cosmological distances. Owing to the progress of optical telescopes, including spaceborne ones, as the JWST, the redshift frontier of known AGN currently exceeds $z=7$ corresponding to the Universe age of 5\% of the present value. The high-redshift objects at this frontier, just as their low-redshift counterparts, appear to harbour and be ``energised'' by supermassive black holes (SMBH). The masses of the latter reach and even exceed $10^{9}M_{\odot}$. In many observable manifestations, the high-redshift AGN appear to be very similar to their low-redshift counterparts. This similarity raises an important question: what are physical processes for which a relatively short time on cosmological scales, hundreds of millions of years or shorter after Big Bang, is sufficient to assemble SMBHs. This topical question is being addressed by many observational techniques in all domains of the electromagnetic spectrum. In radio domain, high angular resolution studies with very long baseline interferometry (VLBI) provide insight into physical processes at the linear scales from hundreds of parsecs down to sub-parsecs throughout the entire range of redshifts. At the high-redshift range, to date these studies have encompassed more than 70 AGN at $z > 4$ (\cite{Krezinger+2022,Gabanyi+2023,Perger+2023PoS} and references therein).

Recently a small sample of four $z > 4.3$ quasars was observed with several VLBI networks in the so called astrometric mode aiming to detect systematic absolute proper motion of cosmological origin (\cite{Titov+2023}). While this study did not produce a conclusive evidence of any anomalies of cosmological origin in the positional variations of these sources, one of them, J2102$+$6015, appeared to deserve attention in view of its potential significance for future multimessenger studies.

 \begin{figure}
    \includegraphics[width=0.96\textwidth]{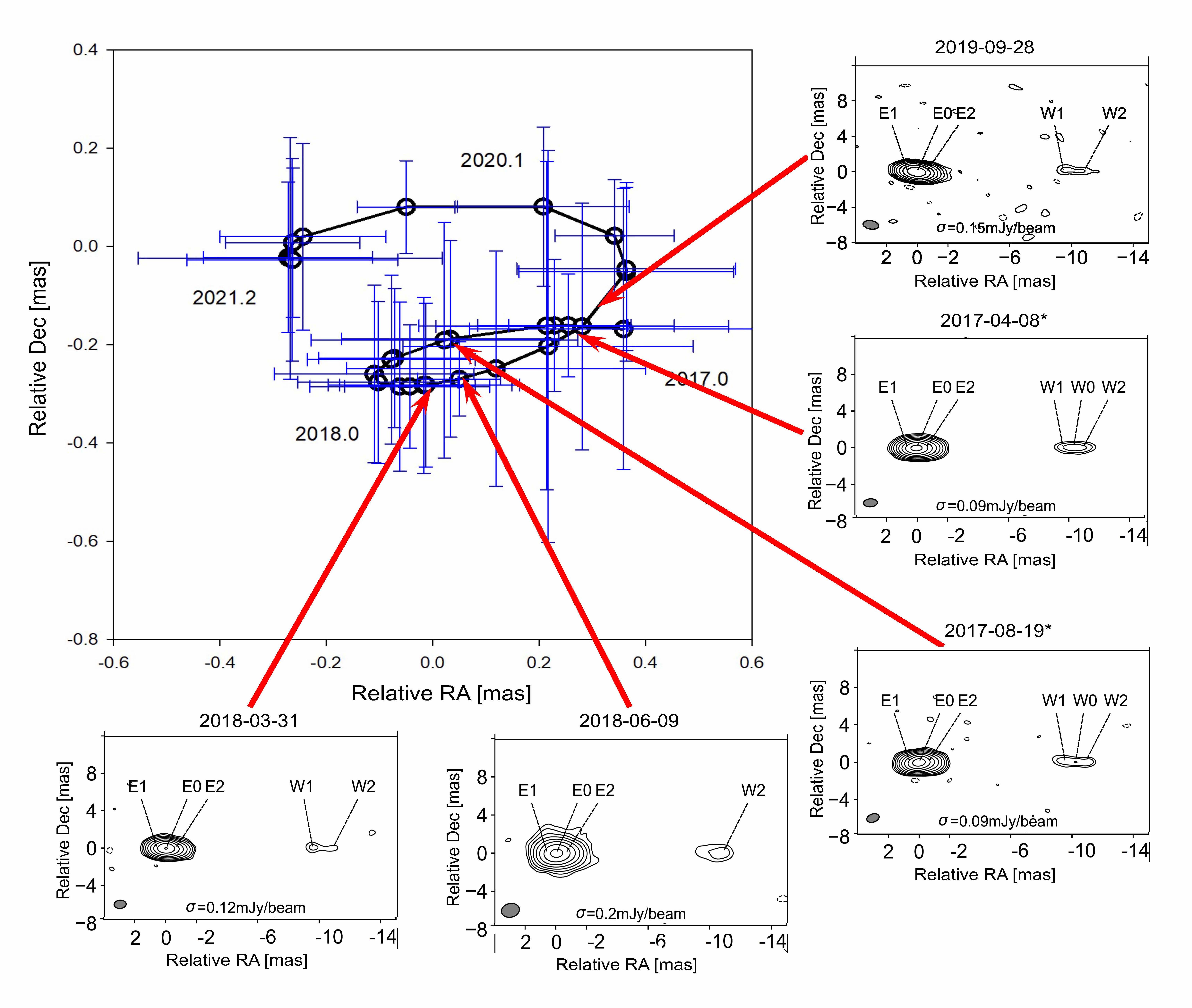}
    \caption{A celestial track of the absolute positions of the source J2102$+$6015 in the period 2017.0 to 2021.2 determined on the basis of VLBI measurements, adopted from \cite{Titov+2023}. The error bars show $\pm 1 \sigma$ formal uncertainty. The insets show the morphology of the source with the total flux density between $\sim 150$~mJy and $\sim 200$~mJy in all epochs obtained in imaging VLBI observations at 8.4~GHz, adopted from \cite{Zhang+2021}. The elliptical Gaussian restoring beam is shown in the lower left corner of each inset. The lowest contours are at $\pm 3 \sigma$ of the image noise shown in each inset, positive contour levels increase by a factor of 2. The images attributed to 2017-04-08 and 2017-08-19 were obtained from VLBI data stacked over several close dates; see \cite{Zhang+2021} for details.}
    \label{f:spiral}
  \end{figure}

\section{An enigmatic object J2102+6015}

The source J2102$+$6015 (2101$+$600) was identified as a quasar at $z=4.5749$; this value was noted by the authors as somewhat unreliable due to a low signal-to-noise ratio of the details in the optical spectrum (\cite{Sowards-Emmerd+2004}). No other measurements of the source's redshift are known to date. The source was observed with various VLBI networks at 2.3, 4.9, 8.4 and 22~GHz (\cite{Frey+2018, Zhang+2021,Frey+2023PoS,Titov+2023}). These observations resulted in the unusual findings requiring further in-depth investigations.

All available to date imaging VLBI results confirm that at all four frequency bands the source's morphology is dominated by two major features separated by about 10~mas in EW direction (see insets in Fig.~\ref{f:spiral}). \cite{Frey+2023PoS} report new European VLBI Network (EVN)  observations at 5 and 22 GHz, and in combination with earlier publications on VLBI results at 2.3 and 8.4~GHz, discuss a set of VLBI images obtained over the period 1994--2021. The eastern feature is stronger and brighter at all frequencies at all epochs. Both features can be model-fitted by two or three Gaussian components. The angular separation between major components of the eastern and western features over the period 2006--2019 is reported to grow with the rate $0.023 \pm 0.011$~mas\,\,yr$^{-1}$, corresponding to the apparent transversal velocity of $(2.8 \pm 1.4)c$ (\cite{Zhang+2022}). However, to date, one and only one observing epoch resulted in a triple structure in J2101$+$6015, in which the known double morphology is amended by a third component located along the same east-west line between the eastern and western features. This is the Very Long Baseline Array (VLBA) observation BZ064 at 8.4~GHz conducted on 2017-02-05. \cite{Zhang+2021} and \cite{Frey+2023PoS} consider arguments for and against classifying the source as a compact symmetric object. Such a classification would be well fitted by the existence of a weak central component associated with the nucleus of the source. However, barring a technical problem in the imaging experiment BZ064 responsible for the single-case erroneous appearance of the central component, a potential variability in the strength and brightness of structural features of the source might have meaningful implications for the model of this object. 

A further intriguing property of the source J2102$+$6015 is reported by \cite{Titov+2023} on the basis of astrometric monitoring over the period 2017--2021 at 2.3 and 8.4 ~GHz. Using a standard for VLBI astrometry method of measuring absolute celestial position of the source, the authors presented apparently quasi-periodic variations of the source's right ascension and declination with the period about 3~years (Fig.~7 in \cite{Titov+2023}) and amplitudes in both coordinates of about 0.8~mas. These variations, if placed on the celestial sphere, produce a quasi-spiral pattern, reproduced in Fig.~\ref{f:spiral}. This pattern enabled the authors to consider a ``toy model'' of the source encompassing a supermassive black hole binary (SMBHB). Indeed, an orbital motion of the binary components might produce apparent oscillations of the celestial coordinates of the source. We note that such an interpretation of the apparent spiral pattern shown in Fig.~\ref{f:spiral} is not unique. However, given the importance of astrophysical and cosmological implications of SMBHB, the model warrants further deliberations and possible verifiable observational manifestations.   

\section{J2102$+$6015, a potential high-redshift multi-messenger}

The ``toy model'' of the source mentioned above (see \cite{Titov+2023}) offers a simple estimate of the semi-major axis of the suspected binary system
\begin{equation}
    a = 0.32\times 10^{-2} \times \left( \frac{M}{10^9 \times M_{\odot}} \right)^{1/3} \times \left( \frac{T\times(1+z)}{3.0 \,\mathrm{yr}} \right)^{2/3} \,\,\, \mathrm{pc},
\end{equation}
\noindent where $a$ is the semi-major axis of the elliptical orbit of the binary system of the total mass $M$ (normalised here to the benchmark value of $10^{9}M_{\odot}$), $T$ is its observer-frame period (in years). Assuming equal masses of $0.5 \times 10^9 M_{\odot}$ for each of the two components for purely illustrative reasons, the semi-major axis of the system would only be $a\simeq 0.25\times 10^{-2}$~pc, or $\simeq 50$ times the gravitational radius of the SMBHB component, $r_\mathrm{g} = 2 G (M/2) c^{-2} \approx 0.48 \times 10^{-4}$~pc.

A potential binary nature of the source J2102$+$6015 has also been suggested by \cite{Zhang+2021}. Indeed, flat radio spectra of the eastern and western morphology features separated by the projected distance of $\sim 70$~pc make them plausible candidates for typical AGN cores. However, their association with components of a suspected SMBHB is inconsistent with the dynamical ``toy model'' described by equation (1): the total mass of such a binary system would be $\sim 10^{16}M_{\odot}$ even without accounting for the projection effect. This is a very unrealistic value. For remaining within the framework of SMBHB nature of the observed oscillation of the absolute celestial position of the source, another binary configuration should be found. The enigmatic central morphological component, reported by \cite{Zhang+2021} might help to find such a configuration. However, given that the astrometric behaviour illustrated in Fig.~\ref{f:spiral} is defined to great extent by the brightest eastern morphological feature, a suspected binary system, responsible for the astrometric oscillations, should be embedded within that eastern feature. 

Irrespective of the specific parameters of the SMBHB model and the source's redshift, the source J2102$+$6015 might be considered as a generator of gravitational waves and a likely contributor into the global gravitational wave background. SMBHB systems evolve inevitably due to the interaction of a less massive object with an accretion disk around a more massive black hole. In this case, the dissipation of the orbital energy occurs, as a result of which the two black holes approach each other (see, for example, \cite{Ivanov+1999}). In the process of decreasing the size of the binary's orbit, the role of gravitational waves increases. Gravitational waves carry away the energy and angular momentum of the binary system (see, for example, \cite{Ivanov+2015}). How does the radiation of  gravitational waves affect the observed parameters of binary systems that are at later stages of evolution than the system considered above, whether such gravitational waves can be detected in the foreseeable future? The source J2102$+$6015 presented here might become an efficient testbed for finding the answers.


\acknowledgements{LIG is grateful to the Leids Kerkhoven-Bosscha Fonds for supporting his attendance at the IAU Symposium No. 375 (grant 22.2.028). SF, K\'{E}G and KP were supported by the Hungarian National Research, Development and Innovation Office (OTKA K134213).} \\ 

\noindent \textbf{Note added in proof:} A recent re-evaluation of the available optical spectrum of the source J2102$+$6015 provides indications that its redshift might be $z=1.24$ rather than $z=4.575$. However, this does not change our view at this source as a potentially highly attractive multimessenger.

\bibliography{IAUS375bib}   

\begin{thebibliography}{}

\bibitem[{Frey} et~al., 2023]{Frey+2023PoS}
{Frey}, S., {An}, T., {Gab{\'a}nyi}, K., {Gurvits}, L., {Krezinger}, M.,
  {Melnikov}, A., {Mohan}, P., {Paragi}, Z., {Perger}, K., {Shu}, F., {Titov},
  O., {de Vicente}, P., \& {Zhang}, Y. 2023, {J2102+6015: an Intriguing
  Radio-loud Active Galactic Nucleus in the Early Universe}.
\newblock {\em PoS}, EVN2022(022), arXiv.2301.07355.

\bibitem[{Frey} et~al., 2018]{Frey+2018}
{Frey}, S., {Titov}, O., {Melnikov}, A.~E., {de Vicente}, P., \& {Shu}, F.
  2018, {High-resolution radio imaging of two luminous quasars beyond redshift
  4.5}.
\newblock {\em A\&A}, 618, A68.

\bibitem[{Gab{\'a}nyi} et~al., 2023]{Gabanyi+2023}
{Gab{\'a}nyi}, K.~{\'E}., {Belladitta}, S., {Frey}, S., {Orosz}, G., {Gurvits},
  L.~I., {Rozgonyi}, K., {An}, T., {Cao}, H., {Paragi}, Z., \& {Perger}, K.
  2023, {Very long baseline interferometry observations of the high-redshift
  blazar candidate J0141-5427}.
\newblock {\em PASA}, 40, e004.

\bibitem[{Ivanov} et~al., 2015]{Ivanov+2015}
{Ivanov}, P.~B., {Papaloizou}, J.~C.~B., {Paardekooper}, S.~J., \& {Polnarev},
  A.~G. 2015, {The evolution of a binary in a retrograde circular orbit
  embedded in an accretion disk}.
\newblock {\em A\&A}, 576, A29.

\bibitem[{Ivanov} et~al., 1999]{Ivanov+1999}
{Ivanov}, P.~B., {Papaloizou}, J.~C.~B., \& {Polnarev}, A.~G. 1999, {The
  evolution of a supermassive binary caused by an accretion disc}.
\newblock {\em MNRAS}, 307(1), 79--90.

\bibitem[{Krezinger} et~al., 2022]{Krezinger+2022}
{Krezinger}, M., {Perger}, K., {Gab{\'a}nyi}, K.~{\'E}., {Frey}, S., {Gurvits},
  L.~I., {Paragi}, Z., {An}, T., {Zhang}, Y., {Cao}, H., \& {Sbarrato}, T.
  2022, {Radio-loud Quasars above Redshift 4: Very Long Baseline Interferometry
  (VLBI) Imaging of an Extended Sample}.
\newblock {\em ApJS}, 260(2), 49.

\bibitem[{Perger} et~al., 2023]{Perger+2023PoS}
{Perger}, K., {Zhang}, Y., {Frey}, S., {An}, T., {Gab{\'a}nyi}, K., {Gurvits},
  L., {Hwang}, C.-Y., {Koptelova}, E., {Paragi}, Z., \& {Wang}, A. 2023,
  {High-resolution Imaging of Two Radio Quasars at the End of the Reionization
  Epoch}.
\newblock {\em PoS}, EVN2022(024), arXiv:2301.09947.

\bibitem[{Sowards-Emmerd} et~al., 2004]{Sowards-Emmerd+2004}
{Sowards-Emmerd}, D., {Romani}, R.~W., {Michelson}, P.~F., \& {Ulvestad}, J.~S.
  2004, {Blazar Counterparts for 3EG Sources at
  -40$^{\circ}<$Decl.$<0^{\circ}$: Pushing South through the Bulge}.
\newblock {\em ApJ}, 609(2), 564--575.

\bibitem[{Titov} et~al., 2023]{Titov+2023}
{Titov}, O., {Frey}, S., {Melnikov}, A., {Shu}, F., {Xia}, B., {Gonzalez}, J.,
  {Tercero}, B., {Gurvits}, L.~I., {de Witt}, A., {McCallum}, J., {Kharinov},
  M., {Zimovsky}, V., \& {Krezinger}, M. 2023, {Astrometric Apparent Motion of
  High-Redshift Radio Sources}.
\newblock {\em AJ}, 165(69), 12.

\bibitem[{Zhang} et~al., 2022]{Zhang+2022}
{Zhang}, Y., {An}, T., {Frey}, S., {Gab{\'a}nyi}, K.~{\'E}., \& {Sotnikova}, Y.
  2022, {Radio Jet Proper-motion Analysis of Nine Distant Quasars above
  Redshift 3.5}.
\newblock {\em ApJ}, 937(1), 19.

\bibitem[{Zhang} et~al., 2021]{Zhang+2021}
{Zhang}, Y., {An}, T., {Frey}, S., {Yang}, X., {Krezinger}, M., {Titov}, O.,
  {Melnikov}, A., {de Vicente}, P., {Shu}, F., \& {Wang}, A. 2021, {J2102+6015:
  a young radio source at z = 4.575}.
\newblock {\em MNRAS}, 507(3), 3736--3744.

\end{thebibliography}

\end{document}